\documentclass{iopart}
\usepackage{comment}

\begin{document}

\title{Casimir-like tunneling-induced electronic forces} 
\author{LM Procopio,$^{1,2}$ C Villarreal,$^3$ and WL
Moch\'an$^1$.}
\address{
$^1$Centro de Ciencias F\'{\i}sicas, Universidad Nacional Aut\'onoma de M\'exico, Apartado Postal 48-3, 62251 Cuernavaca, Morelos, M\'exico.}
\address{
$^2$Facultad de Ciencias, Universidad Aut\'onoma del Estado
de Morelos, Avenida Universidad 1001, 62221 Cuernavaca, Morelos, M\'exico.}
\address{
$^3$Instituto de F\'{\i}sica, Universidad Nacional 
Aut\'onoma
de M\'exico, Apartado Postal 20-364, 01000 Distrito Federal, M\'exico.}

\eads{\mailto{LMPP <lmpp@fis.unam.mx>}, \mailto{CV
<carlos@fisica.unam.mx>}, \mailto{WLM 
<mochan@fis.unam.mx>}} 

\begin{abstract}
We study the quantum forces that act  between two nearby
conductors due to electronic tunneling. We derive an expression for
these forces by 
calculating the flux of momentum  
arising from the overlap of evanescent electronic
fields. Our result is written in terms of the electronic reflection
amplitudes of the conductors and it has the same structure as
Lifshitz's formula for the electromagnetically 
mediated Casimir forces.
We evaluate the tunneling force between two semiinfinite conductors
and between two thin films separated by
an insulating gap. We discuss some applications of our results.
\end{abstract} 


\submitto{\JPA}

\section{Introduction}

The increased accuracy of
experimental studies \cite{lamoreux,mohideen,capasso,decca,onofrio} of
the Casimir force 
\cite{casimir} 
between conducting bodies has opened the possibility 
of exploring new ideas related to the understanding and control of
quantum vacuum fluctuations. Research projects on
the dynamical Casimir effect \cite{dynamical}, Casimir torques \cite{torques},  
or the possible
applications of the Casimir forces in the development of micro and
nano-electromechanical devices are now under way \cite{serry,esquivel}.
Understandig  Casimir forces has become fundamental in the investigation of
deviations of Newton's Gravitational Law at micrometer dimensions,
related to the search for extra dimensions in space-time \cite{gravity}.
 
The usual Casimir effect may be understood as a force due to the quantum
nature of the electromagnetic radiation.
In this paper we study another source of quantum forces, namely, the
tunnel effect. Particles that are able to tunnel across a barrier have
more space available to them. Thus, their contribution to
the total energy of a many body system such as two metallic slabs with 
neighboring surfaces may diminish. As the amount of space gained and
the number of 
particles capable of tunneling depends on the width of the barrier,
besides its 
height, there must be a force that performs work when the width is
modified. 
Since quantum tunneling arises from evanescent electronic fields,
this force is similar to the contributions of   
evanescent electromagnetic waves to the standard Casimir force.

In this paper we concentrate our attention on conduction electrons in
conductors, that is, on massive non-relativistic Fermions. 
We derive the tunneling force by calculating the flux of momentum 
between two regions delimited by an arbitrary potential $V(x)$.
We express the momentum flux in terms of the Green's function
of the system, which we evaluate by means of a scattering method
involving amplitude reflection coefficients \cite{lambrecht,mochan}. 
This method yields an expression for the tunneling force with a
structure that is
essentially identical to Lifshitz's formula \cite{lifshitz}. We first
perform the calculation 
for a one-dimensional system. We then extend the calculation to
the three-dimensional case. Finally, we evaluate the tunneling
force for a configuration consisting of two semiinfinite or two thin
metallic slabs separated by a thin insulating gap and  we discuss some
applications of our results.

\section{One dimensional systems}

The dynamical equation for  the wavefunctions of an electronic
system may be derived from a
Lagrangian density, \cite{goldstein}
\begin{equation}
\mathcal L = \frac{\hbar^2}{2m} \vert\psi_{,z}\vert^2+V\vert\psi\vert^2 
+ \frac{i\hbar}{2} \left(\psi\psi_{,t}^* -
\psi^*\psi_{,t})\right),
\end{equation}
for which Euler-Lagrange's equations yields Schr\"odinger's equation, 
\begin{equation} 
\partial_t \frac{\partial\mathcal{L}}{\partial(\psi_{,t}^*)}+
\partial_z\frac{\partial\mathcal{L}}{\partial(\psi_{,z}^*)}-
\frac{\partial\mathcal{L}}{\partial\psi^*}
= i\hbar \psi_{,t}+ \frac{\hbar^2}{2m}\psi_{,z,z} -V\psi=0.
\end{equation}
The wavefunction carries mechanical properties which may be derived
from $\mathcal{L}$: we may obtain a momentum density
\begin{equation} \label{momentum}
g= \frac{1}{c}T^0_{z}=\frac{\partial \mathcal L}{\partial\psi^*_{,t}}\psi^*_{,z} 
+ \frac{\partial\mathcal{L}}{\partial\psi_{,t}}\psi_{,z} 
=\frac{i\hbar}{2}(\psi\psi^*_{,z}-\psi^*\psi_{,z}),
\end{equation}
where $c$ is the speed of light in vacuum, as well as a momentum flux \cite{bogoliubov}
\begin{equation} \label{flux}
\fl \qquad
 T^z_z=\frac{\partial \mathcal{L}}{\partial\psi^*_{,z}}\psi^*_{,z} +
\frac{\partial \mathcal{L}}{\partial\psi_{,z}}\psi_{,z}
-\mathcal{L}
=
\frac{\hbar^2}{2m}
\vert{\psi_{,z}}\vert^2-V\vert\psi\vert^2-\frac{i\hbar}{2} 
\left(\psi \psi^*_{,t}- \psi^* \psi_{,t}\right).
\end{equation}

Consider now an eigenstate $\psi_n$ of the Schr\"odinger Hamiltonian
$\hat H=-(\hbar^2/2m)\partial_z^2+\hat V$ corresponding to an energy
$E_n$.  From 
Eq. (\ref{flux}), its contribution to the momentum flux is
\begin{equation} \label{tn}
T^z_z=\frac{\hbar^2}{2m}\vert\partial_z\psi_n\vert^2
+(E_n-V)\vert\psi_n\vert^2. 
\end{equation}
Within a region $\mathcal V$ in which $V(z)$ may be taken as a constant, we can
write $E_n-V=\hbar^2 k^2_n/2m$, i.e., the kinetic energy of particles
with wavenumber $k$ within $\mathcal V$. We now sum the contributions
(\ref{tn})  over all the occupied orbitals,
\begin{equation} \label{flux2}
T^z_z(z)=\frac{\hbar^2}{2m}\sum_{n} \int dE\, \delta(E-E_n)f(E)
\left[k^2\vert\psi_n(z)\vert^2 +\vert\partial_z\psi_n(z)\vert^2\right],
\end{equation}
where $f(E_n)$ is the occupation number of orbital $n$, given in
equilibrium by the Fermi-Dirac distribution function, and
$k^2=2m(E-V)/\hbar^2$. 
The energy integration and Dirac's $\delta$ allow us to write $T_z^z$ 
in terms of the Green's
function of the system, 
\begin{equation} \label{green1}
\hat{G}_{E}(z,z')=\left\langle
z\left|\left(E-\hat{H}\right)^{-1}\right|z'\right\rangle =
\sum_n\frac{\psi_n(z)\psi_n^*(z')}{E-E_n},
\end{equation}
employing the relation $\mbox{Im} 
(E^+ - E_n)^{-1}=-\pi \delta(E-E_n)$, 
where $E^+=E+i\eta$ with $E$ and $\eta\to0^+$ real. 
Substituting this latter relation in (\ref{flux2}) and employing
(\ref{green1}) we obtain
\begin{equation} \label{flux3}
\fl\qquad
T^z_z(z)=-\frac{\hbar^2}{2\pi m} \mbox{Im} \int dE\left[k^2G_{E^+}(z,z')
+\partial_z \partial_{z'} G_{E^+}(z,z')\right]_{z'\to z}f(E).
\end{equation}
Notice that Eq.(\ref{flux3}) may be interpreted as
\begin{equation} \label{rhoef} 
T^z_z(z)=\int dE \rho^{ef}_E f(E)(\hbar k)\left(\frac{\hbar k}{m}\right),
\end{equation}
where  $\pm\hbar k$ is the momentum of a particle which moves at
velocity $\pm\hbar k/m$, 
thus contributing the amount $(\hbar k)\hbar k/m$ to the momentum
flux, and 
\begin{equation} 
\rho^{ef}_E(z)=-\frac{1}{2\pi} \mbox{Im}\left[G_E(z,z')+
\frac{1}{k^2} \partial_z\partial_{z'} G_E(z,z')\right]_{z'\to z}
\end{equation}
plays the role of an effective local density of states.

We now assume that $\mathcal V$ has a width $L$ and is bounded on both
sides by arbitrary potentials, and we evaluate the Green's function
following a scattering approach \cite{lambrecht,mochan}.  Within $\mathcal V$, the
solution of $(E-H)G_E(z,z')=\delta(z-z')$  may be written as 
$ G_E(z,z')= ({2m}/{\hbar^2}){\psi_L(z_L) \psi_R(z_R)}/{W}$, 
where $\psi_L$ and $\psi_R$ are the two solutions of the
Schr\"odinger-like homogeneous 
equation $(E-H)\psi=0$ that satisfy the boundary conditions on the left and the
right side of the system respectively, $W =
\psi_L \psi_{R,z}-\psi_{L,z} \psi_R$ is their Wronskian, and
$z_L$ and $z_R$ are the smallest and the largest among $z$ and $z'$. 
We write
$ 
\psi_{L}(z)= e^{-ikz_L} + r_1 e^{ikz_L}
$
and
$
\psi_{R}(z)= e^{ik(z_R-L)} + r_2 e^{-ik(z_R-L)},
$
where $r_1$ and $r_2$ are the reflection amplitudes for particles
impinging on the left and right boundaries of
$\mathcal V$, which we assume at $z=0$ and $z=L$, and 
we 
obtain
\begin{equation}
G_E(z,z')=\frac{2m}{\hbar^2}
\frac{\left(e^{-ikz_L} + r_1 e^{ikz_L}\right)\left(e^{ik(z_R-L} + r_2 e^{-ik(z_R-L}\right)}
{2 ik e^{-ikL}\left(1-r_1r_2e^{2ikL}\right)},
\end{equation}
which together with Eq.(\ref{flux3}) yields the momentum flowing
within $\mathcal V$,
\begin{equation} \label{stress1}
T^z_z= \frac{1}{\pi} \mbox{Re} \int dE\,  k
\frac{1+r_1r_2e^{2ikL}}{1-r_1r_2e^{2ikL}} f(E), 
\end{equation}
which may be conveniently written as
\begin{equation} \label{stress2}
T^z_z= \frac{\hbar^2}{\pi m} \mbox{Re} \int dk\, k^2 
\frac{1+r_1r_2e^{2ikL}}{1-r_1r_2e^{2ikL}} f(E)
\end{equation}
by using the relationship $E=\hbar^2 k^2/2m+V$ and changing integration 
variable. Notice that, as expected in an equilibrium situation,
$T^z_z$ is independent of $z$.

\section{Three dimensional systems}

The generalization of the results derived above to the three
dimensional case is straightforward for 
systems which are translationally invariant along a symmetry plane,
say $xy$. In that case, the parallel wave vector $\vec Q=(Q_x,Q_y)$
is a conserved quantity, and for each $\vec Q$ the
problem is identical to 
the 1D case.
Thus, we
only have to sum Eq. (\ref{stress1}) over the allowed wavevectors,
\begin{equation} \label{67}
\mathcal T^z_z=  \frac{\hbar^2}{4\pi^3m}
\mbox{Re} \int d^2Q\,\int dk\, k^2
\frac{1+r_1r_2e^{2ikL}}{1-r_1r_2e^{2ikL}} f(E)
\end{equation}
where we introduced the number $\mathcal{A} d^2Q/(2\pi)^2$  of 
wavevectors within a region $d^2 Q$ of reciprocal space by applying
Born-von Karman boundary conditions in a system
with total area $\mathcal{A}\to\infty$, and we introduced the momentum
flux {\em density} $\mathcal T_z^z=T_z^z/\mathcal A$. Notice that
$-\mathcal T^z_z$ coincides with the $zz$ component of the stress
tensor as defined in elasticity theory. 

We remark that the structure of Eq. (\ref{67}) is essentially
identical to Lifshitz's formula for the Casimir effect between two
materials when written in terms of their optical reflection
coefficients \cite{lambrecht,mochan}. The main differences are that 
the electromagnetic field has
two independent transverse polarizations whose contributions would
have to be summed over, and that the speed of light is a constant $c$,
while the speed of electrons is proportional to the wavevector, i.e.,
the dispersion relation between electrons and photons are different,
and consequently, there is an extra power of $\hbar$ in
Eq. (\ref{67}).

As $r_1$ and $r_2$ are independent of $\vec Q$ for
scalar fields, the first integral in Eq. (\ref{67}) may be performed
immediately. At zero temperature we obtain 
\begin{equation} \label{70}
\mathcal T_z^z=\frac{1}{\pi^2} \mbox{Re} \int dk\,
\left(K_F-\frac{\hbar^2}{2m}k^2\right)k^2 
\frac{1+r_1r_2e^{2ikL}}{1-r_1r_2e^{2ikL}}, 
\end{equation}
where the integration region includes all states below the Fermi
level, whose kinetic energy within $\mathcal V$ is $K_F$, and for which we
took $f(E)=2$, including the spin degeneracy.

\section{Applications}

\subsection{One semiinfinite metal}
Within the bulk of a semiinfinite metal the electrons are reflected by
the surface potential 
barrier on one side, while there is no barrier on the other
side. Thus, the pressure $p$ with which the electrons push the surface of
the metal may be obtained by setting $r_1=0$ in Eq. (\ref{70}). The
result is simply
\begin{equation} \label{71}
p=\mathcal T_z^z=\frac{1}{\pi^2}\int^{k_F}_0 dk
\left(E_F-\frac{\hbar^2 k^2}{2m}\right)k^2
= \frac{2}{5}nE_F,
\end{equation}
where $\hbar k_F$ is the Fermi momentum, $E_F=K_F$ (within the metal)
is the Fermi 
energy, and  $n=k^3_F/3\pi^2$ is the electronic density. As could have
been expected, this result coincides with the well known pressure of
a degenerate fermion gas \cite{pathria}.

\subsection{Two semiinfinite metals}

We consider now two identical semiinfinite metals separated by
vacuum. The force 
$F/\mathcal A$ per unit area between both metals may be obtained from the 
momentum flux (\ref{70}) within the vacuum region, where the
wavefunction of all the occupied states are evanescent, and it may
be written as 
\begin{equation}\label{Tzzresta} 
\frac{F}{\mathcal A}=-2\mbox{Im}\frac{\hbar^2}{2m \pi^2}\int_{\kappa_0}^{\kappa_F}
d\kappa\, (\kappa_F^2-\kappa^2) \kappa^2 \frac{1}{\zeta-1},
\end{equation} 
where $\zeta^{-1}=r^2 e^{-2 \kappa L}$, and 
we wrote the wavenumber $k=i\kappa$ in terms of the decay constant
$\kappa$. 
 The integration limits in (\ref{Tzzresta}) are  the decay constants
for electrons at the bottom of the conduction band, $\kappa_0=\sqrt{[}2
m (W+E_F)/\hbar^2]$, and at the Fermi level $\kappa_F=\sqrt(2 m
  W/\hbar^2)$, while $W=K_F$ (within vacuum) is the work function, and
$r=r_1=r_2$ is the  
complex reflection 
amplitude corresponding to evanescent wavefunctions that {\em
propagate} (i.e., decay) through vacuum towards a surface and are reflected
back. Assuming that the potential $V(z)$ is constant within the metals
and within vacuum, and that it changes abruptly at the vacuum-metal interface
by an  amount $W+E_F$, the reflection amplitude may be calculated as
$r=(i\kappa-k_M)/(i\kappa+k_M)$,
where $k_M=\sqrt{[}2m(W+E_F)/\hbar^2-\kappa^2]$ is the wavenumber within
the metal of the state corresponding to $\kappa$.

\begin{figure}
\centering{\input{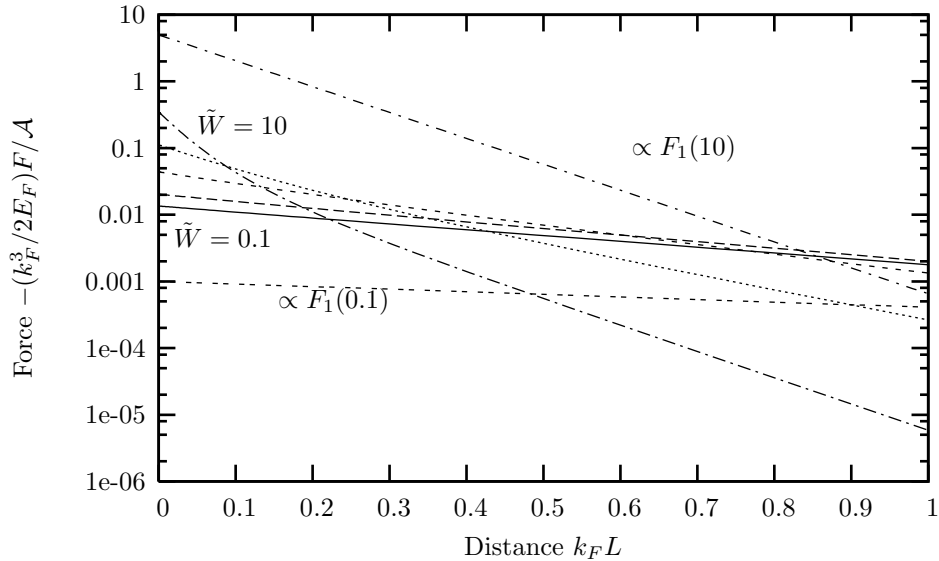}}
\caption{\label{FvsPhi}Force por unit area between two semiinfinite metals as
function of their separation $L$ for different values $0.1, 0.3,1,3,10$
of the dimensionless work function $\tilde W=W/2E_F$. Also shown
schematically is 
the contribution $e^{-2\kappa_F L}\propto F_1$ expected for one
electron at the Fermi surface in the cases $\tilde W=0.1,10$.}
\end{figure}

In Fig. \ref{FvsPhi} we plot the force per unit area as a function of
distance for different values of the workfunction $W$.
The force is attractive, seems to decay exponentially for {\em large}
separations and attains a finite value at zero separation. 
For large $W$  the force is larger at small separation and smaller at
large separations as the energy decays very fast towards that of two
isolated semiinfinite metals.
As could have been expected, the
smaller the work function, the larger the spatial range of the
force. We might 
expect the decay to be dominated by those electrons closest to the
Fermi energy whose contribution becomes proportional to 
$e^{-2\kappa_F L}$. Fig. \ref{FvsPhi} includes two curves illustrating
this behavior for the cases of large and small $W$.  The actual
decay of the force
is slightly faster, more so for small $W$. This is due to the fact
that not only the contribution of each electron decays with increasing
distance, but
also the number of electrons that contribute effectively to the
force. Furthermore, the phase space available right at the Fermi
energy is null, due to the prefactor 
$\kappa_F^2-\kappa^2$ in Eq. (\ref{Tzzresta}), so the contributing
electrons have a slightly larger decay constant (i.e., smaller range)
than those at the Fermi level.

\begin{figure}
\centering{\input{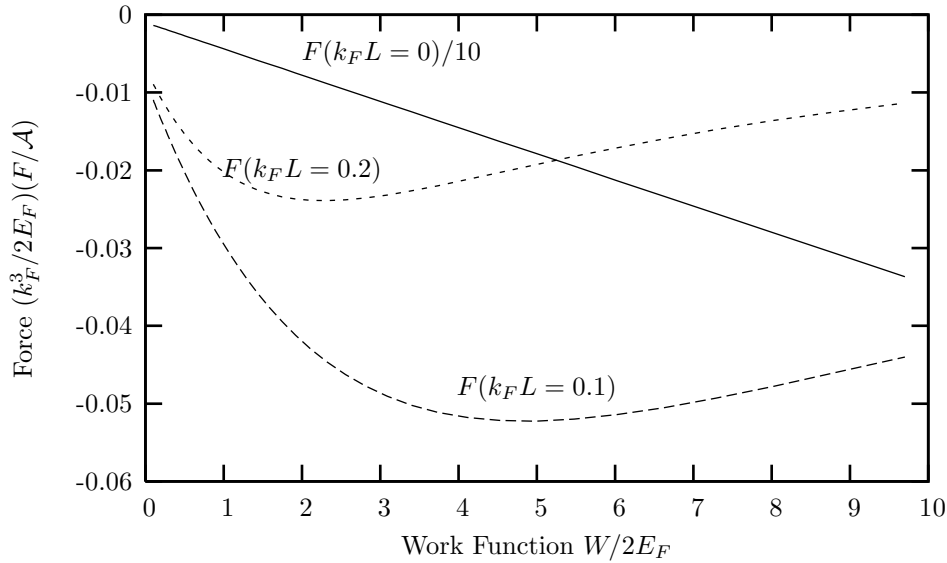}}
\caption{\label{FvsPhiL0}Force between two metals as a function of the
work
function for several values of the separation $\tilde L=0,0.1,0.2$ in
units of $k_F^{-1}$.} 
\end{figure}

In Fig. \ref{FvsPhiL0} we show the force for several distances as a
function of $W$. 
For finite separation distances the force is small when $W$ is large,
as the surfaces don't {\em feel} each other anymore, and it is 
also small for small workfunction, as the electrons tunnel too easily
between the two metals, not {\em caring} about the separation. Thus,
the magnitude of the force is largest at 
some intermediate value of $W$ which increases as $L$ decreases. At
contact, $L=0$, there is no such extreme anymore and we obtain a
linear behavior,
\begin{equation}\label{Tzz0b}
\frac{F(0)}{\mathcal
A}=-\frac{q_F^3}{\pi^2}\left(\frac{E_F}{5}+\frac{W}{3}\right),
\end{equation}
as can be shown by integrating Eq. (\ref{Tzzresta}) analytically.
$F(0)$ is the force that would be required in order to break an infinite
metal into two semiinfinite ones. Eq. (\ref{Tzz0b}) actually overestimates the
ultimate breaking strength of real materials by several orders of
magnitude as our model fails to account for dislocations whose motion
within the metal would relax the stress, and for the growth of fractures
which are actually responsible for the failure of real metals. Real
metals break gradually, not simultaneously over the whole separation
surface. Nevertheless, integrating Eq. (\ref{Tzzresta}) over $L$
we have obtained an analytical estimate of the surface
energy of metals in terms only of their Fermi energy and their work
function. This turns out to be surprisingly
accurate \cite{prl} given our simplifying assumptions, namely, our use of an
independent free  particle model, neglecting the crystalline structure,
the electronic charge, and many body corrections, as well as our use
of a square potential barrier at the surface.

\section{Thin films}

\begin{figure}
\centering{\input{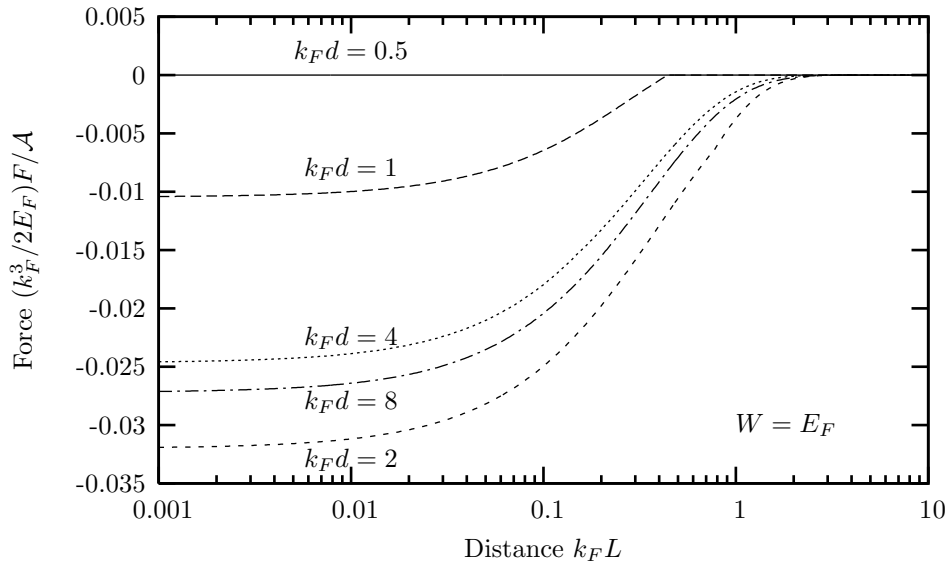}}
\caption{\label{fFvsL}Force between two films of widths $\tilde
d=0.5,1,2,4,8$ (in units of 
$k_F^{-1}$) as a function of the distance $L$ between them. We took
$W=E_F$.}
\end{figure}

Eq. (\ref{Tzzresta}) may be employed to calculate the force between more
complicated systems simply by introducing the
appropriate value of the reflection amplitude. For example, in Fig. 
\ref{fFvsL} we display the force between two free standing very thin
metallic films as a function of distance for a fixed Fermi energy, or
more properly, a fixed electrochemical potential. 
Notice that for very thin films the force is identically zero, as
there are no states at all below the Fermi energy 
and therefore there are no available electrons to tunnel between the
films. For wider films the force is finite at small separations but
becomes zero after a finite separation. As the width is further
increased, the force
approaches that corresponding to semiinfinite metals, although not
monotonously; it actually oscillates between larger and smaller values. 
This behavior may be understood by considering the finite size effect
on the levels of the individual films, and the interaction of levels
within both films, yielding alternating bonding and antibonding states
which may be occupied only when they lie below the Fermi level.

\section{Conclusions}
By calculating the mechanical properties carried by the electronic
wavefunctions, we have shown that the interchange of electrons between
conductors produces a force that may be calculated in terms of the
electronic reflection amplitudes using formulae that are very closely
related to Lifshitz formula for the usual Casimir force. We
illustrated our formalism by calculating the electronic pressure
within a conductor and the force between semiinfinite conductors and
between thin films at very small distances, of the order of the Fermi
wavelength. These distances are extremely small, beyond the expected
limit of validity of the usual Casimir effect. Thus, we expect our
results to be important to study the forces that act, for example,
between the tip and the substrate of a scanning tunneling microscope 
\cite{experiment2}.
We discussed how our results may be employed to calculate the surface
energy of conductors without having to substract total energies. Other
applications which are currently under study include the calculation
of the force between impurities embedded within three and
one-dimensional conductors. Although  our current calculations were
performed  for free, independent electron conductors, we believe that
our scattering approach might be generalized to more realistic systems
of interacting electrons. 

\section*{Acknowledgment}
We acknowledge useful discussions with H. Larralde, R. Esquivel,
D. Iannuzzi, and U. Mohideen. This work was
partially supported by DGAPA-UNAM under grants No. IN117402 and 
No. IN118605.

\end{document}